# Ab initio phonon transport across grain boundaries in graphene using machine learning based on small dataset

Amirreza Hashemi[1], Ruiqiang Guo[2,#], Keivan Esfarjani[3,4,5], and Sangyeop Lee[2,6*]

[1] Computational Modeling and Simulation Program,
University of Pittsburgh, Pittsburgh PA 15261

[2] Department of Mechanical Engineering and Materials Science,
University of Pittsburgh, Pittsburgh PA 15261

[3] Department of Mechanical and Aerospace Engineering,
University of Virginia, Charlottesville VA 22904

[4] Department of Materials Science and Engineering,
University of Virginia, Charlottesville VA 22904

[5] Department of Physics,
University of Virginia, Charlottesville VA 22904

[6] Department of Physics and Astronomy,
University of Pittsburgh, Pittsburgh PA 15261

[#]current affiliation: Thermal Science Research Center, Shandong Institute of Advanced Technology, Jinan, Shandong Province, 250103, China.

* sylee@pitt.edu

## Abstract

Establishing the structure-property relationship for grain boundaries (GBs) is critical for developing next generation functional materials, but has been severely hampered due to its extremely large configurational space. Atomistic simulations with low computational cost and high predictive power are strongly desirable, but the conventional simulations using empirical interatomic potentials and density functional theory suffer from the lack of predictive power and high computational cost, respectively. A machine learning interatomic potential (MLIP) recently emerged but often requires an extensive size of the training dataset, making it a less feasible approach. Here we demonstrate that an MLIP trained with a rationally designed small training dataset can predict thermal transport across GBs in graphene with *ab initio* accuracy at an affordable computational cost. In particular, we employed a rational approach based on the structural unit model to find a small set of GBs that can represent the entire configurational space and thus can serve as a cost-effective training dataset for the MLIP. Only 5 GBs were found to be enough to represent the entire configurational space of graphene GBs. Using the atomistic Green's function approach and the MLIP, we revealed that the structure-thermal resistance relation in graphene does not follow the common understanding that large dislocation density causes larger thermal resistance. In fact, thermal resistance is nearly independent of dislocation density at room temperature and is higher when the dislocation density is small at sub-room temperature. We explain this intriguing behavior with the buckling near a GB causing a strong scattering of flexural phonon modes. Our work shows that a machine learning technique combined with conventional wisdom (e.g., structural unit model) can extend the recent success of *ab initio* thermal transport simulation, which has been mostly limited to single crystals, to complex yet practically important polycrystals with GBs.

## Introduction

Grain boundaries (GBs) are of interest in many applications because they are common defects and largely affect electrical, mechanical, and thermal properties. For two dimensional (2D) materials such as graphene, experimental studies showed that GBs commonly exist in graphene sheets prepared by exfoliation[1-5], causing the fundamental physical properties of polycrystal samples largely deviate from those of single crystals. Therefore, engineering GBs is an effective way to achieving desired electronic, thermal, and mechanical properties in many applications[6-14].

The physical properties are largely dependent on the local atomic structure of GB[5,6,15] and thus it is important to establish the structure-property relationship on how a GB structure affects the physical properties. However, establishing such a structure-property relationship has been challenging mainly for two reasons. The first is that GBs have extremely large configurational space. For example, three dimensional (3D) materials have 5 degrees of freedom (misorientation angle noted as $\theta_M$ hereafter, line angle, and three degrees of freedom of crystalline grain orientation in 3D space) for GB structures, making the configurational space extremely large. The second is that the experimental characterization of individual GB requires significant efforts particularly for preparing samples with a geometrically well-defined GB. The samples with GBs have been prepared by bonding two wafers with a twist angle but it often leaves void at the interface[16]. Therefore, it is challenging to experimentally study enough number of GBs to draw a statistically conclusive finding on the structure-property relationship.

Atomistic simulation can be a useful tool for the study of GBs if it has high predictive power, but also has major challenges. The atomistic simulation for thermal transport such as molecular dynamics (MD)[7,8,10,17-19] and the atomistic Green's function (AGF)[14,20] require an interatomic potential. A common approach for the interatomic potential has been empirical potentials that have a rigid functional form parametrized based on quantum mechanical calculation results and experimental data. Although the empirical potentials have been useful for promoting the understanding of physical phenomena from an atomistic level, they have clear limitations. For the physical properties that were not considered for the parametrization, empirical potentials do not provide an accurate prediction. Also, because of its rigid functional form, it is usually not flexible enough to describe a wide range of atomic configurations. On the contrary, *ab initio* calculation can be highly accurate and have a predictive power without adjustable parameters as demonstrated by the recent studies. For example, the high thermal conductivity of boron arsenide was experimentally confirmed[21-23] after the prediction from *ab initio* simulation[24]. Also, the significant hydrodynamic phonon transport in graphitic materials was predicted using *ab initio* simulation first[25,26] and then experimentally confirmed[27,28]. However, the *ab initio* simulation for thermal transport has been limited to single crystalline phase and point defect cases. For the thermal transport across GBs, the *ab initio* simulation is not feasible due to its high computational cost considering the size of GB atomic structures.

A recently emerging method is to use machine learning schemes to predict the interatomic interactions based on the dataset from *ab initio* simulations[29-40]. This so-called machine learning interatomic potential (MLIP) was motivated by the fact that the interatomic interaction is a function in a high dimensional space where machine learning outperforms conventional regression methods. Recently developed MLIPs show that the MLIP can be as accurate as *ab initio* calculations while its computational cost is several orders-of-magnitude cheaper than the *ab initio* calculations[29-31,41]. In particular, the MLIP was proven for predicting the thermal transport in crystalline phase[29,31,34] and partially disordered crystalline phase that has vacancies[29]. This confirms that the MLIP is accurate enough to correctly capture subtle anharmonicity, which is critical for phonon-phonon scattering and phonon-strain field scattering, and is also flexible enough to describe various atomic configurations including vacancies. However, extending the past success of MLIP to spatially extended disorder case (e.g., GBs) has some challenges. Unlike vacancies, the GBs have extremely large atomic configurational space. Therefore, the training dataset should be carefully designed such that it can represent the entire configurational space. In addition, the size of the training dataset should be minimal since generating the training dataset from *ab initio* calculation can be prohibitively expensive considering the typical size of GB structures.

In this work, we develop MLIPs using the Gaussian regression, called the Gaussian approximation potential (GAP) [41,42], for studying phonon transport across graphene GBs. We use a systematic framework based on the structural unit model to select the complete and orthogonal training dataset. With the carefully chosen a few GBs for the training dataset, we show that the GAP can produce identical results as the *ab initio* calculations for the wide range of GBs while its computational cost is 6 orders of magnitude cheaper than the *ab initio* calculations. Using the GAP and AGF, we then report several important features of phonon transport across GBs in graphene with its high predictive power. We distinguish the influence of dislocation core and extended strain field on phonon scattering, and reveal an intriguing scattering of flexural phonon modes by out-of-plane buckling in graphene GBs. We also briefly evaluate an empirical potential (Tersoff) that has been widely used in past studies by comparing it to GAP.

## Results

*Identifying the small set of GBs representing the entire configurational space of GBs*

A challenge in developing an MLIP for GBs is how to prepare a complete set of training data. Considering the typical period length of GBs and the area strained by a GB, a supercell that contains a GB can be too large for the *ab initio* calculation. Thus, for the training dataset, it is critical to select a small set of GBs that can represent the entire configurational space of GBs. In early studies developing an MLIP for general purpose, a fraction of the total database was chosen for the training dataset without much rationale, with the remaining as the testing dataset[29,41-44]. Recently, active learning schemes have been proposed to reduce the size of training datatset[45-47], making it possible to simulate the dynamic evolution of systems such as phase change in a large scale for a long time period. While the active learning scheme can be used for general cases, it does not allow to use of preexisting knowledge on the system of study even when it is available. Besides, the active learning scheme is more suitable for molecular dynamics simulation in which a training dataset is added based on the measured uncertainty at each time step. For phonon transport simulation, the lattice dynamics-based method (e.g., AGF) has several important advantages over molecular dynamics simulations such as modal analysis and no statistical error.

We use the fact that most GBs have hierarchical structures with basic building blocks as demonstrated in the previous studies that analyzed the GB structures with the structural unit model[48-50]. A basic idea is to identify those basic building blocks or *unique* local atomic environments (LAEs) from many GBs and find a small set of GBs that contain the complete set of the unique LAEs[51]. Then, an MLIP trained with the data from the small set of GBs is expected to accurately capture the interatomic interactions of GBs in the entire configurational space. We analyzed 20 GBs covering the full span of $\theta_M$ which contains a total of 5544 LAEs. In this work, we focus on symmetric GBs with zero line angle because several parameters that are expected to affect phonon scattering such as GB formation energy, dislocation density, and out-of-plane roughness are nearly unchanged with the line angle in graphene[4]. The LAEs were described using the smooth overlap of atomic position (SOAP) descriptors[52], which show the significant overlap among the 5544 LAEs. By comparing the similarity of the SOAP vectors from the total 5544 LAEs[51], we could identify that the total 5544 LAEs can be reduced to only 12 and 13 unique LAEs for the structures relaxed by Tersoff empirical potential (TSF)[53,54] and density functional theory (DFT), respectively. The TSF and DFT produce slightly different structures after relaxation, and hence the number of unique LAEs also differ. The analysis shows that the total 20 GBs covering the full span of $\theta_M$ can be composed using those 12 or 13 unique LAEs, confirming the idea that the extremely large configurational space of GBs in fact have a very small number of basic building blocks. We then identified 5 representative GBs shown in Figure 1 that contain all of the 12 or 13 unique LAEs. The selected GBs significantly differ from each other in terms of the topological arrangements and the density of disclinations. We used the 5 GBs to generate a training dataset for our GAP, train the GAP, and performed the AGF simulation with the GAP to simulate the phonon transport across GBs as discussed in the method sections.

*Validation of the simulation framework using TSF*

We use the TSF potential to validate our simulation framework from selecting representative GBs to the AGF calculation. Unlike the *ab initio* calculation, the TSF potential is computationally cheap enough to generate the data of interatomic force constants and transmission function of all the 20 GBs. Therefore, the GAP trained with the TSF data (called GAPTSF hereafter) can be directly validated against the results from TSF for all the 20 GBs. In Figure 2, we compare the GAPTSF and TSF for the GB formation energy, and spectral phonon transmission function. The GAPTSF and TSF agree well with each other for the prediction of the GB energy for both the training and testing GBs. In particular, the spectral phonon transmission functions, the property of interest in this work, are identical for all GBs. This confirms that the 5 GBs chosen from the LAE analysis are enough to represent the entire 20 GBs and thus the resulting GAP is highly accurate and reliable for a wide range of GBs.

*Results from GAP trained with datasets from DFT (GAPDFT)*

With the success of GAPTSF, we proceeded to developing GAPDFT using the training dataset from density functional theory (DFT) calculation. Like GAPTSF, the GAPDFT also shows excellent accuracy. The root-mean-square of errors (RMSE) of energy and force are 0.0011 eV and 0.052 eV/Å respectively for the training set, and the RMSE of energy and force are 0.0019 eV and 0.066 eV/Å respectively for the testing set. In Figure 3, we examine the GAPDFT compared to DFT for the relaxed atomistic structures. The structures relaxed by the GAPDFT are identical to those by DFT in particular for the out-of-plane atomic displacements.

Figure 4 presents the GB formation energy from GAPDFT and DFT, showing good agreement between them for the entire range of $\theta_M$. The overall trend of GB formation energy from the GAPDFT follows the trend predicted by the Read-Shockley model[55]; the GB formation energy is linear to $\theta_M$ for low $\theta_M$ (<15°) and high $\theta_M$ (>45°) while the mid-range $\theta_M$ show non-monotonic behavior of GB formation energy with respect to $\theta_M$.

In Figures 4b and 4c, we separate the GB formation energy into the contribution from local dislocation cores (called core energy, $E_{\text{core}}$) and surrounding strain field (called strain energy, $E_{\text{strain}}$)[56,57] to better understand the GB formation energy and its effects on phonon transport. We should note that this is one of the noteworthy advantages of MLIPs. The MLIPs can predict each atom's contribution to total energy while DFT cannot in principle. The core energy ($E_{\text{core}}$) and strain energy ($E_{\text{strain}}$) can be defined as:

$$E_{\text{core}} = \frac{\sum_i^{N_{\text{core}}} E_i - \frac{N_{\text{core}}}{N_{\text{tot}}} E_{\text{bulk}}}{l_{\text{unit}}} \quad (1)$$

$$E_{\text{strain}} = \frac{\sum_i^{N_{\text{strain}}} E_i - \frac{N_{\text{strain}}}{N_{\text{tot}}} E_{\text{bulk}}}{l_{\text{unit}}} \quad (2)$$

where $N_{\text{core}}$ and $N_{\text{strain}}$ are the number of atoms forming dislocation cores (pentagons and heptagons) and hexagon lattices, respectively. The $N_{\text{tot}}$ is the total number of atoms. The $E_{\text{bulk}}$ and $l_{\text{unit}}$ are the energy per atom in the perfect crystalline phase and the length of GB. The core energy and strain energy from GAPDFT in Figures 4b and 4c seem physically reasonable. The dislocation density linearly increases with $\theta_M$, have a maximum value at mid-$\theta_M$, and linearly decreases with $\theta_M$ (see Figure S3). Therefore, the core energy in Figure 4b is maximum in the mid-$\theta_M$ range where the dislocation density is maximum. The strain energy is minimum in the same $\theta_M$ range where the lattice can open up to insert one additional lattice plane to form an edge dislocation and thus the strain is minimized[58].

In Figure 5, we present the thermal resistance as a function of $\theta_M$ at various temperatures from the AGF and the Landauer formalism calculations. At high temperatures of 500 K and 1500 K in Figures 5c and 5d, the thermal resistance has a concave shape with respect to $\theta_M$, having a maximum resistance value at mid $\theta_M$ range. This behavior is similar to the case of Si and diamond at 1000 K that a previous study reports using molecular dynamics simulation with an empirical potential[8]. A common explanation for this behavior has been that the dislocation density is the maximum in the mid-$\theta_M$ and thus the phonon scattering by GBs is expected to be maximum in the mid-$\theta_M$ range. However, we observe different behaviors at low temperatures at 300 K and 100 K. At 300 K in Figure 5b, the concave shape of thermal resistance becomes negligible and the resistance is nearly independent of the $\theta_M$. As temperature further decreases to 100 K in Figure 5a, the thermal resistance shows a convex shape with respect to $\theta_M$, having the lowest thermal resistance at mid-$\theta_M$. The behavior of thermal resistance at 100 K and 300 K is clearly opposite to the current understanding that the higher dislocation density leads to higher thermal resistance. For graphene GBs, the higher dislocation density does not necessarily lead to higher thermal resistance. In particular, at 100 K, the thermal resistance is even higher when the dislocation density is smaller.

A possible explanation for this intriguing behavior of thermal resistance as a function of $\theta_M$ at different temperatures is that dislocation core and nearby strain field affect the phonon scattering by GBs to the different extents at different temperatures. At low temperatures, heat is mostly carried by long wavelength phonons which experience only weak scattering by dislocation cores since the wavelength is much longer than the characteristic size of the cores. The strain field can be a major contributor to the phonons scattering at low temperature due to its spatially extended characteristics. This is supported by the fact that the strain energy distribution in Figure 4c and the thermal resistance at 100 K in Figure 5a have a similar trend with respect to $\theta_M$; both thermal resistance and strain energy are minimum in the mid-$\theta_M$. At high temperatures where the short wavelength phonons are the major heat carriers, the wavelengths become comparable to the size of dislocation cores which thus cause strong scattering due to its nature of large lattice distortion compared to the strain field. The thermal resistance at 500 K and 1500 K in Figure 5 follow a similar trend as the core energy in Figure 4b.

Observing the important role of the strain field for phonon scattering at low temperatures, we further investigate its detailed mechanisms. Figures 6a and 6b show the thermal conductance normalized by the ballistic thermal conductance of perfect graphene as a function of temperature. The normalization eliminates the specific heat effects from the conductance and thus shows how much the thermal conductance is suppressed by phonon scattering at a GB at various temperatures. The total 20 GBs can be clearly separated into two groups: one showing monotonously decreasing normalized thermal conductance as a function of temperature shown in Figure 6a and the other showing increasing at low temperature and then decreasing normalized thermal conductance with temperature shown in Figure 6b. It is interesting to see that most GBs of the first group are from mid-$\theta_M$ while the latter group is from the small and large $\theta_M$. To explain the different behavior of the two GB groups, we consider spectral transmissivity defined as the phonon transmission function across a GB normalized by the ballistic phonon transmission function across single crystalline graphene. In Figure 6c, we present the spectral transmissivity for the two GBs with $\theta_M$ of 6.02° and 32.20° that represent each group. In the frequency range below 15 THz which dominates the thermal transport below room temperatures, the two GBs show a remarkable difference. While the spectral transmissivity is high and nearly constant for the GB with $\theta_M$=32.20°, the transmissivity for the GB with $\theta_M$=6.02° is low and increases rapidly with frequency. It is noteworthy that the majority of phonon states below 15 THz are from the flexural acoustic phonon branch due to the quadratic phonon dispersion and large density-of-states.

The remarkably different scattering of flexural modes in the two GB groups is originated from the structural difference, in particular buckling induced by a GB. This is consistent with the previous studies[59,60] that showed flexural modes are strongly scattered by buckling of GB structure. Figure 6d shows that the two groups of GBs are very different in terms of out-of-plane buckling. The common disclinations in graphene, pentagon and heptagon, create compression and dilation stress at the tips of disclinations, respectively. When a GB has low or high $\theta_M$, the pentagon and heptagon disclinations are far from each other due to the low density of dislocations, and thus the out-of-plane buckling is induced to reduce the compressive and dilation strain. On the contrary, when a GB has a mid $\theta_M$, the disclination cores are densely packed along the GB line with the pentagon and heptagon cores placed next to each other. In such a case, the compressive and dilation strain are canceled and the out-of-plane buckling does not occur[55]. Therefore, at low temperatures where the thermal phonon wavelength is comparable to the characteristic length of buckling, the significant buckling in GBs with low and high $\theta_M$ causes strong scattering of the flexural phonon modes. As a result, the GBs with low and high $\theta_M$ exhibit higher thermal resistance at 100 K than those with mid $\theta_M$ in Figure 5a, although they have lower dislocation density.

*Comparison of TSF and GAPDFT*

Lastly, it would be interesting to present a brief comparison of GAPDFT and TSF since the TSF has been widely used in past studies while its accuracy for phonon transport across GBs has not been comprehensively examined. In Figure 4, we compare GAPDFT and TSF for the GB formation, core, and strain energies. Figure 4a shows that the TSF overestimates the GB formation energy compared to the GAPDFT. This is because the core energy from TSF is larger than that from GAPDFT in the mid-$\theta_M$ range where the density of dislocation core is maximum as shown in Figure 4b. On the contrary, for strain energy in Figure 4c, the TSF and GAPDFT show similar predictions for the wide range of $\theta_M$ although the strain energy from TSF is slightly smaller. The comparison of the core and strain energy from TSF and GAPDFT indicates that TSF is reasonably accurate in predicting the energy of strained hexagon structure while poor in predicting the energy of severely distorted structures such as pentagons and heptagons.

The thermal resistances from TSF and GAPDFT in Figure 5 are observed similar, but the force constants and spectral transmission functions behind the thermal resistance values are noticeably different for TSF and GAPDFT. For the self-interaction force constant in the crystalline phase, the TSF overpredicts by 35% compared to the GAPDFT (see Supplementary Information S5). The force constant prediction by TSF has more pronounced error in the core region of GBs. In Figures 7a and 7b, we present the error of TSF in predicting force constant change upon the introduction of GBs. We define the normalized error as $|\Delta\Phi_{ii,\mathrm{TSF}}-\Delta\Phi_{ii,\mathrm{GAPDFT}}|/\Delta\Phi_{ii,\mathrm{GAPDFT}}$ where $\Phi_{ii}$ is a self-interaction force constant and $\Delta\Phi_{ii}$ is the difference of a self-interatomic force constant from the perfect crystalline case (i.e., $\Phi_{ii,\mathrm{GB}}-\Phi_{ii,\mathrm{crystal}}$). The figure shows that the error in the core region is pronounced and reaches up to 50% while the error is small for the surrounding hexagons. This agrees with the aforementioned observation that the TSF has significant error for dislocations while is reasonably accurate for strained hexagons. As a result, the spectral transmissions from GAPDFT and TSF in Figures 7c and 7d show substantial difference above 20 THz where dislocation cores are important for phonon scattering. Overall, the suppression of transmission functions from the perfect crystalline phase is noticeably larger in TSF than in GAPDFT, also supported by the overprediction of core energy by TSF in Figure 4b. However, below 20 THz where the strain field is the dominant cause for phonon scattering, the GAPDFT and TSF show similar suppression of the spectral transmission function.

## Discussion

In summary, we demonstrated that MLIPs trained with the rationally designed minimal dataset can predict phonon transport across GBs with *ab initio* predictive power and accuracy while the computational cost is affordable. A special attention was paid on reducing the required training dataset by employing the idea of structural unit model that GBs have hierarchical structures and have only a few basic building blocks. Our approach shows that only 5 GBs are enough to represent the entire configurational space and thus the small training dataset using those 5 GBs is sufficient for an MLIP. Indeed, our test using TSF and GAPTSF shows that force constants and spectral transmission functions from the TSF and GAPTSF are nearly identical for 20 GBs covering the entire configurational space.

The GAPDFT trained with the dataset from DFT reveals several intriguing characteristics of phonon scattering by GBs with *ab initio* accuracy. Previous studies for three dimensional bulk materials suggested that thermal resistance increases with dislocation density, but we showed that graphene does not follow the same trend. The thermal resistance at room temperature does not depend on the dislocation density and even decreases with increasing dislocation density. We explained this with the two dimensional structural characteristics of graphene: flexural phonon modes carrying the majority of heat and out-of-plane buckling induced by GBs. The heat-carrying flexural phonon modes are strongly scattered by the out-of-plane buckling which is pronounced for the GBs with low dislocation density. Thus, dislocation density alone cannot determine the scattering of phonons in polycrystalline graphene but the surrounding strain field plays an important role.

We also briefly examined the accuracy of TSF for thermal transport across GBs by comparing it to GAPDFT. The overall thermal resistance values from both TSF and GAPDFT reasonably agree with each other, but the force constants and spectral transmission functions show a noticeable difference. In particular, TSF shows inaccuracy in predicting dislocation cores (pentagons and heptagons) while is reasonably accurate for the strain field. As a result, the transmission functions from TSF agree with those from GAPDFT at low frequency where the strain field is important for phonon scattering, but shows noticeable error in the mid to high frequency range.

Our work provides deep insights into the atomic-level mechanisms governing phonon transport across graphene GBs, particularly for the buckling effects on phonon transmission and thermal resistance. This understanding may help to explain phonon transport across GBs in other two-dimensional materials and also to engineer their thermal properties using GBs. The present method for developing MLIPs with minimal training dataset can be easily extended to three dimensional materials. It would help to predict and understand thermal transport in the polycrystalline phase of emerging materials for which a reliable interatomic potential has not been developed yet.

## Methods

*Finding a small set of representative GBs*

The common disclinations in graphene are pentagon and heptagon rings and each LAE includes a few rings of pentagon, hexagon, and heptagon. We selected 20 GBs that covers the full span of $\theta_M$ (0° to 60°) which include a variety of disclination densities and different topological arrangement of disclinations. The $\theta_M$ and coincidence site lattice (CSL) $\Sigma$ values of the 20 GBs are listed in table S1 of the supplementary information (SI).

We generated 20 supercells that contain those 20 GBs. For structure relaxation using MD simulations, a periodic boundary condition is preferred for all directions of the supercells. However, a GB breaks the translational symmetry along the direction perpendicular to the GB line. We, therefore, make a supercell that includes two GBs of the same type with the opposite GB direction such that the supercell is periodic along the direction perpendicular to the GB line. To generate such a structure, we first constructed a supercell that contains a single GB using an algorithm based on the centroidal Voronoi tessellation[4]. Then, we appended the same supercell that is rotated by 180° and the resulting supercell is periodic along the direction perpendicular to the GB line. We then relaxed the obtained supercell, particularly for buckling, by running MD simulations at 300 K in the NVT ensemble using the LAMMPS package, with a time step of 0.5 fs and TSF potential. For the DFT calculations, we further optimized the structure of supercells relaxed by MD simulations using the energy minimization scheme in the VASP package. The DFT calculations were performed using ultrasoft pseudopotentials with a plane wave cutoff energy of 286 eV. The convergence criteria for energy and force were set to $10^{-8}$ eV and $10^{-6}$ eV/Å, respectively.

We used the SOAP[52] descriptor to find the smallest GB dataset that contains all the representative LAEs in the 20 GBs. The SOAP descriptor places a Gaussian function on each atom to construct the density of neighbor atoms $\rho_i$, which is then expanded in a basis set of radial functions $g_n(r)$ and spherical harmonics $Y_{lm}(\mathbf{r})$ as

$$\rho_i(\mathbf{r}) = \sum_{nlm} c_{nlm}^{(i)} g_n(r)\, Y_{lm}(\mathbf{r}),$$

where $c_{nlm}^{(i)}$ are the expansion coefficients for atom *i*. The descriptor is formed from these coefficients by computing the power spectrum elements

$$p_{nn'l}^{(i)} = \frac{1}{\sqrt{2l+1}} \sum_m c_{nlm}^{(i)} \left(c_{n'lm}^{(i)}\right)^*.$$

The resulting descriptor has invariance under translation, rotation, and the permutation of atoms. For each GB, a SOAP descriptor for each atom $i$ in the GB is calculated and represented as coefficients of basis functions $\boldsymbol{p}_i = \{p_1, p_2, \cdots, p_N\}$. The length of the SOAP vector $N$ is determined by a radial basis cutoff $n_{\max}$ and an angular basis (spherical harmonic) cutoff $l_{\max}$. We evaluate the dissimilarity of LAEs using SOAP descriptors which is defined as[51]:

$$d_{ij} = \sqrt{\boldsymbol{p}_i \cdot \boldsymbol{p}_i + \boldsymbol{p}_j \cdot \boldsymbol{p}_j - 2\boldsymbol{p}_i \cdot \boldsymbol{p}_j}$$

where $\boldsymbol{p}_i$ and $\boldsymbol{p}_j$ are the SOAP vectors for the two atoms $i$ and $j$. We introduce a parameter ε, serving as a criteria for the unique LAE. If $d_{ij} > \varepsilon$, the $\boldsymbol{p}_i$ and $\boldsymbol{p}_j$ are different from each other indicating that the two atoms $i$ and $j$ are surrounded by different LAEs. Otherwise, we determine $\boldsymbol{p}_i$ and $\boldsymbol{p}_j$ represent the same LAE. In this work, we used 0.04 for the value of ε.

*Training GAP*

We performed MD simulations of the 5 representative GBs to generate training data, which are the snapshots of the atomic position, force, and configurational energy. The MD simulations were performed at 300 K in the NVT ensemble with a time step of 0.5 fs. After initial time steps for thermal equilibration, we took one snapshot every 50 time steps to reduce the correlation between snapshots. The training datasets for both GAPTSF and GAPDFT include relaxed structures of the 5 selected GB structures and 50 snapshots for each GB at 300K. After obtaining the training dataset, we used the same hyperparameters to fit the GAP models, as shown in Table S2 of the Supplementary Information.

*AGF simulation*

For the AGF simulation, the supercell needs to be sufficiently large so that the leads do not have strain from a GB. The supercell we used for the AGF calculation is 10 times longer in the direction perpendicular to GBs than those we used for training GAP. Since the AGF simulation does not require translational symmetry along the heat flow direction, the supercells for the AGF calculation contain only one GB for each unlike those for training the GAP that have two GBs. See Supplementary Information S3 for the details of supercells. We used phonopy[61] and LAMMPS[62] to calculate second-order force constants. In the AGF simulation, we used decimation technique[63,64] to approximate surface Green's functions and we used a frequency broadening factor of 1 $cm^{-1}$ for the continuous representation of discrete eigenfrequencies. We observed a good convergence of transmission function with 20 transverse wavevectors for the GB with the largest width ($\theta_M$=50.57°). For other GBs, the number of transverse wavevectors was determined such that the product of the number of transverse wavevectors and the width of GB is the same for all GBs.

## Acknowledgments

A.H. and S.L. acknowledge support from National Science Foundation (Award No. 1943807). The simulation of this work was performed using the Linux cluster of the Center for Research Computing at the University of Pittsburgh.

## Author Contributions

S.L. conceived the idea and supervised the research. R.G. and A.H. trained interatomic potentials. A.H. and K.E. performed AGF calculation. S.L. and A.H. analyzed the AGF results. All authors wrote and commented on the manuscript.

## Competing Interests

The authors declare no competing financial interests.

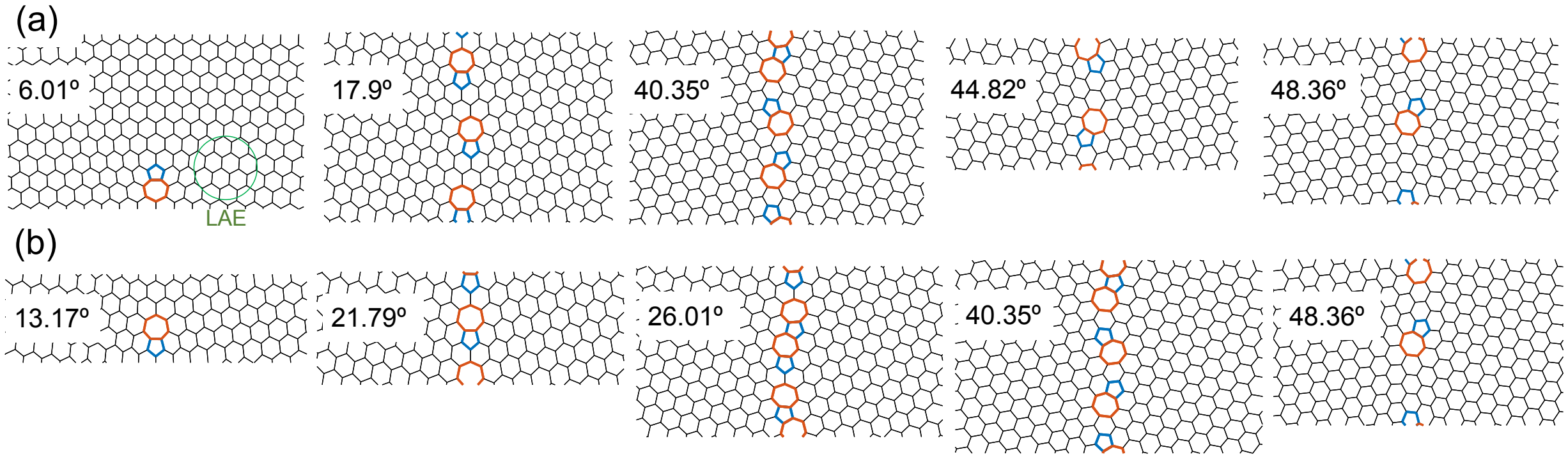

**Figure 1**. Five representative GBs from (a) TSF and (b) DFT showing distinct features such as density of disclinations and the their topological arrangements. The angle in each figure shows the misorientation angle. The green circle shows the cutoff radius for defining LAE.

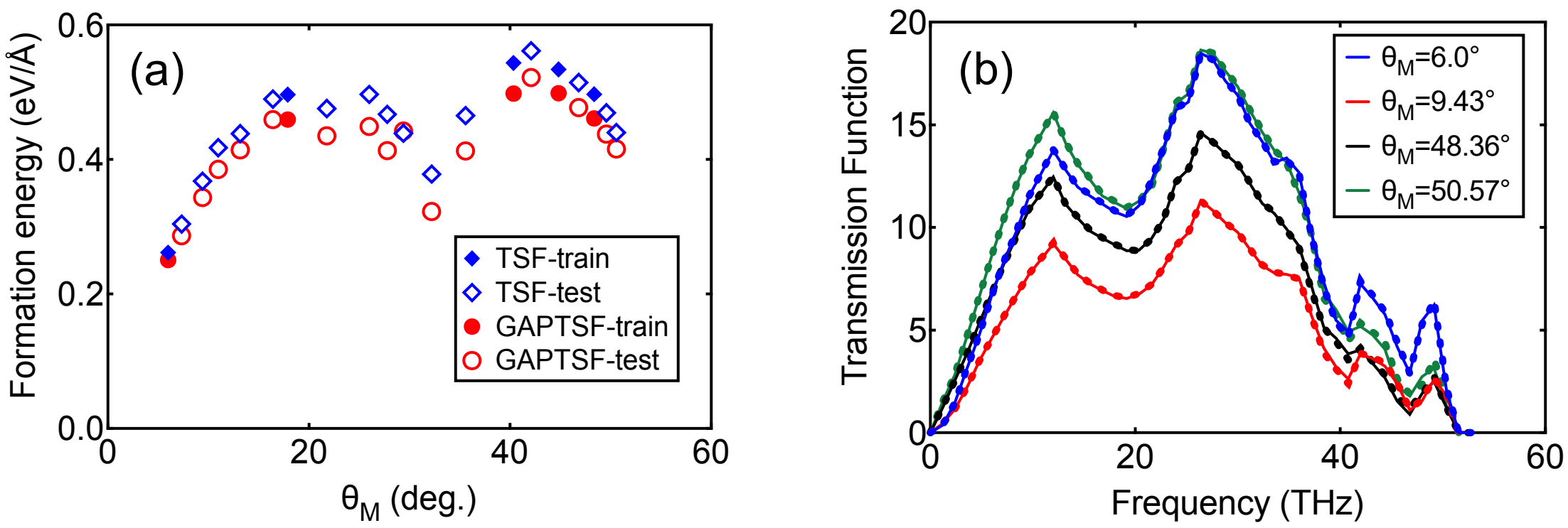

**Figure 2**. Validation of GAPTSF against TSF for (a) formation energy of GBs, and (b) transmission function. The solid symbols in (a) represent GBs used for training the GAPTSF. The solid lines and dots in (b) are from GAPTSF and TSF, respectively. In (b), the two GBs with $\theta_M$=6.0° and 48.36° and the other two GBs with $\theta_M$=9.43° and 50.57° are from the training and testing dataset, respectively.

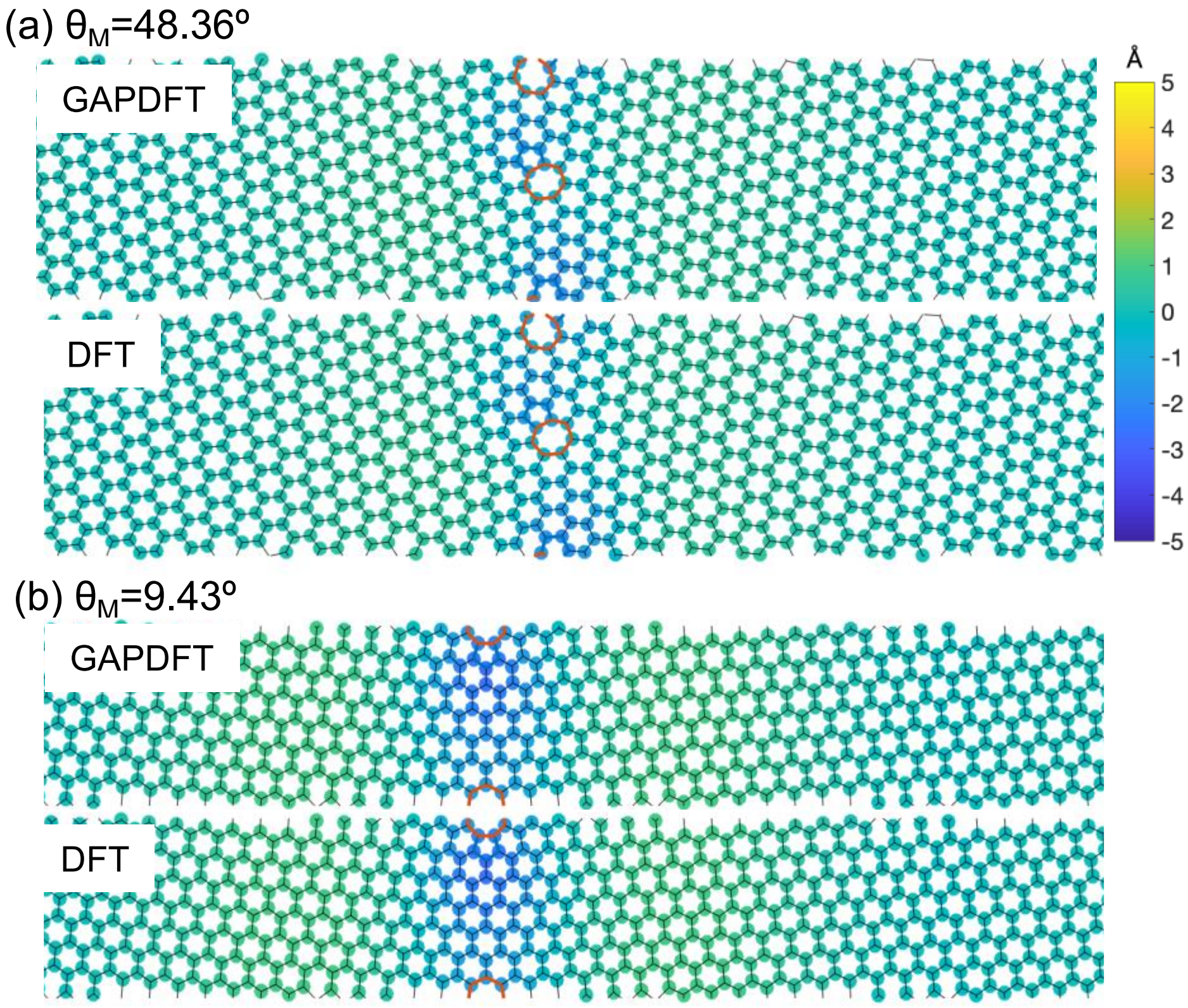

**Figure 3**. Validation of GAPDFT against DFT for relaxed structures projected onto a-b plane. (a) $\theta_M$=48.36º from the training dataset and (b) $\theta_M$=9.43º from the test dataset. The color represents out-of-plane displacement in Å.

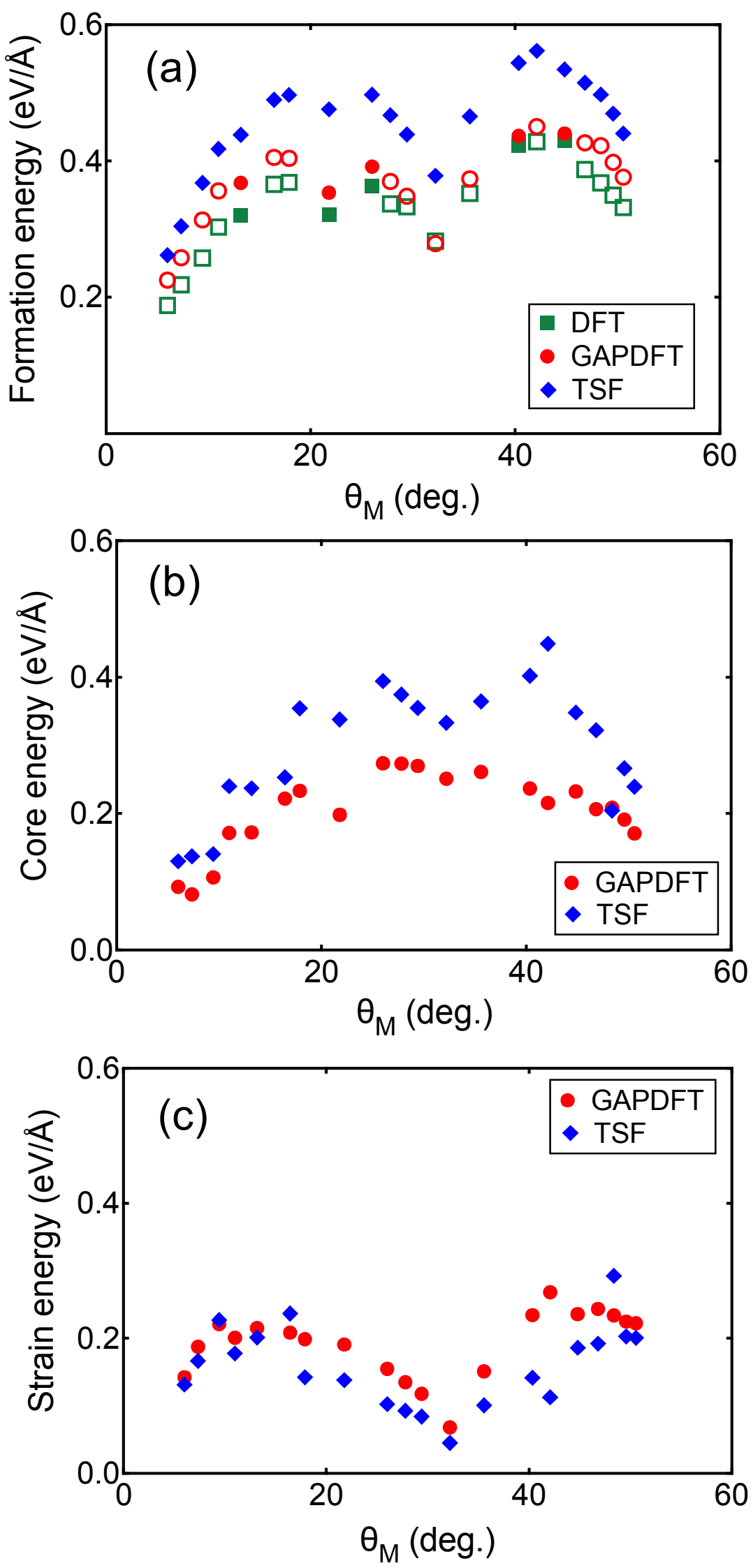

**Figure 4**. Comparison of DFT, GAPDFT, and TSF for (a) GB formation energy, (b) core energy, and (c) strain energy. The solid symbols in (a) represent the GBs that were used for training GAPDFT.

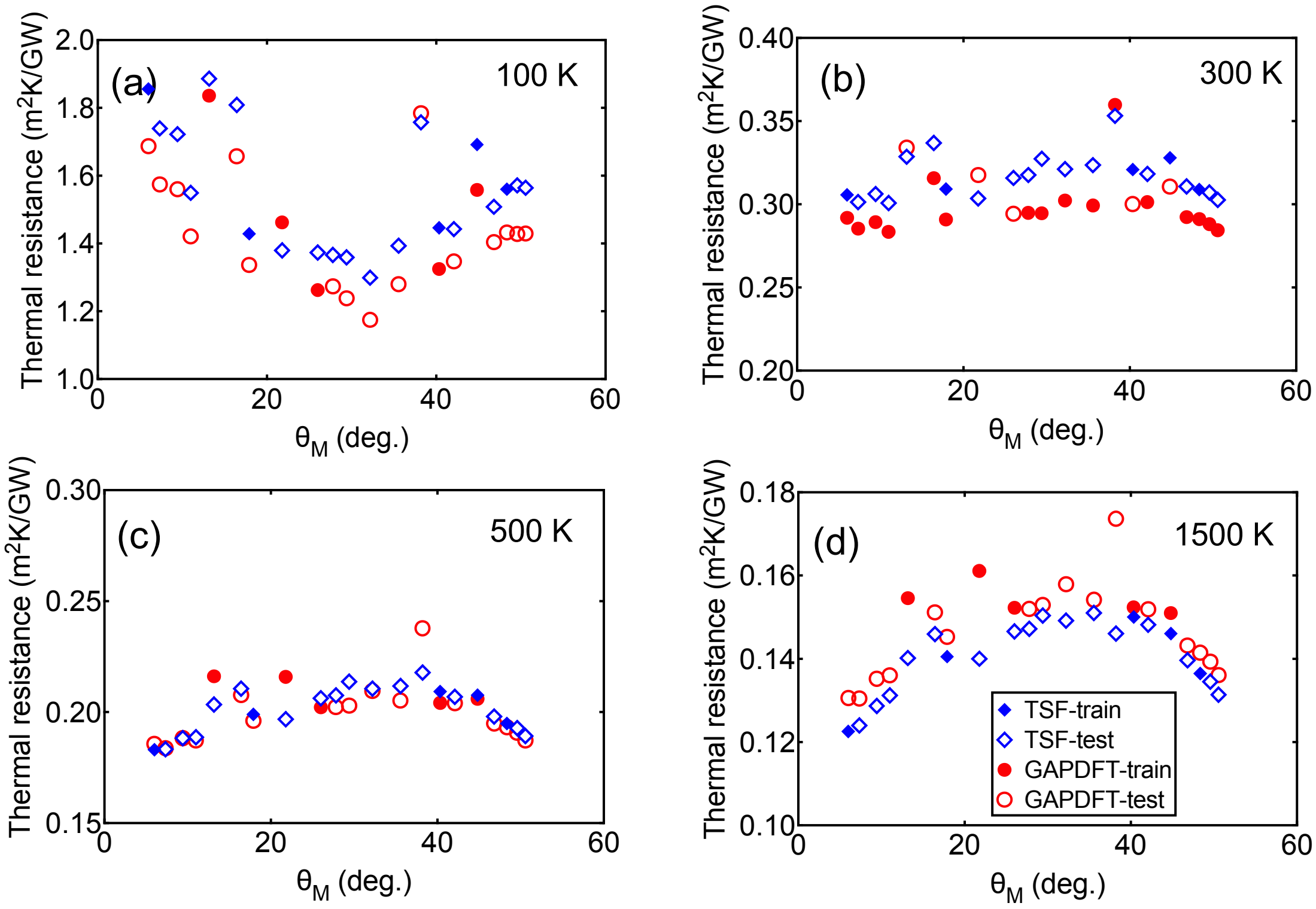

**Figure 5**. Thermal resistance with varying misorientation angles at (a) 100 K, (b) 300 K, (c) 500 K, and (d) 1500 K.

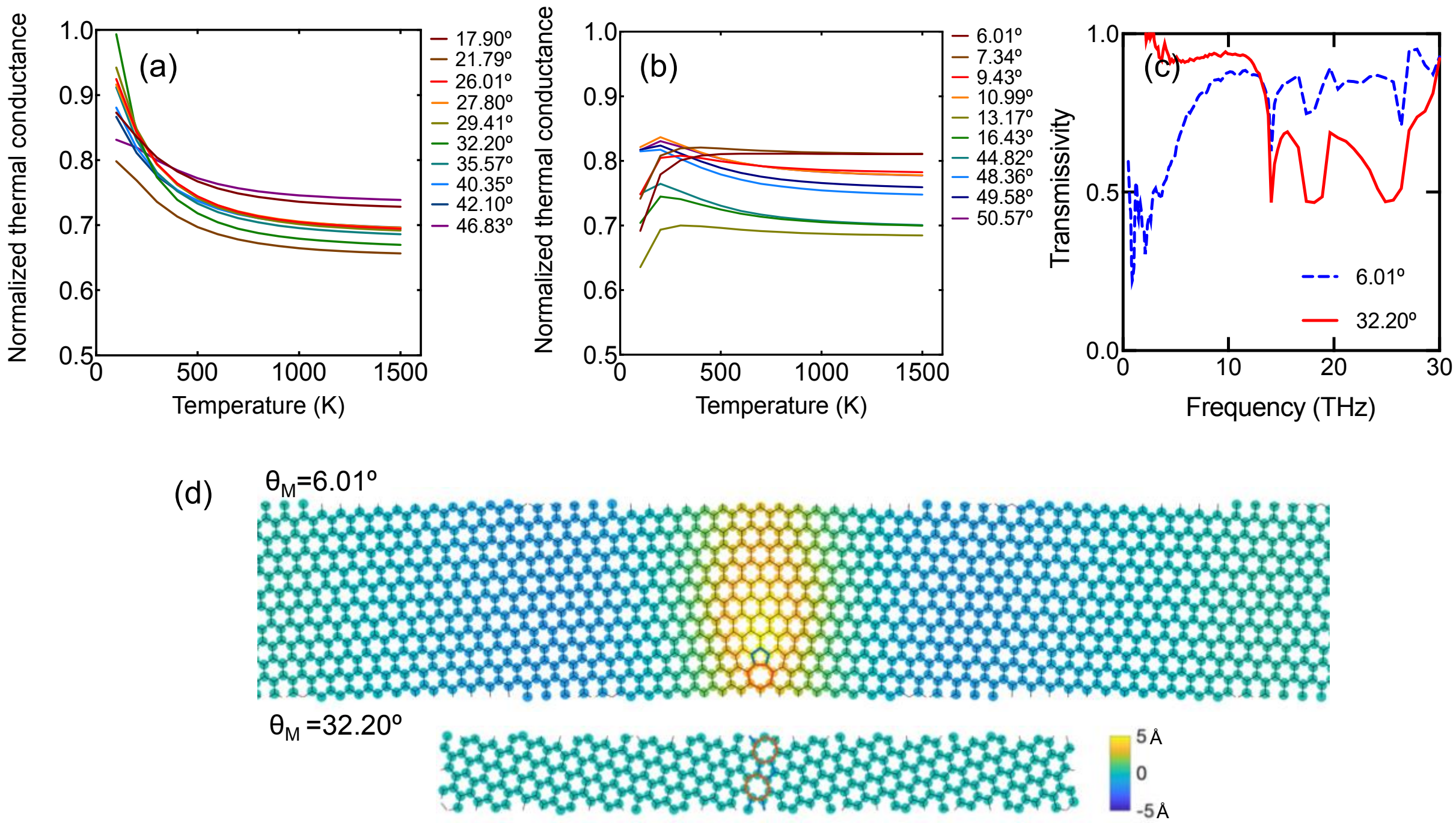

**Figure 6.** Role of out-of-plane buckling for scattering of flexural phonon modes. (a-b) normalized thermal conductance as a function of temperature for (a) GBs showing monotonously decreasing behavior and (b) GBs showing increasing behavior at low temperatures. The values in the legends represent misorientation angle. (c) Phonon transmissivity for two representative GBs showing a remarkable difference in low phonon frequency range below 15 THz. (d) Comparison of the two representative GBs in terms of out-of-plane buckling. The color represents out-of-plane displacement of atoms and the pentagon and heptagon are marked in blue and red, respectively.

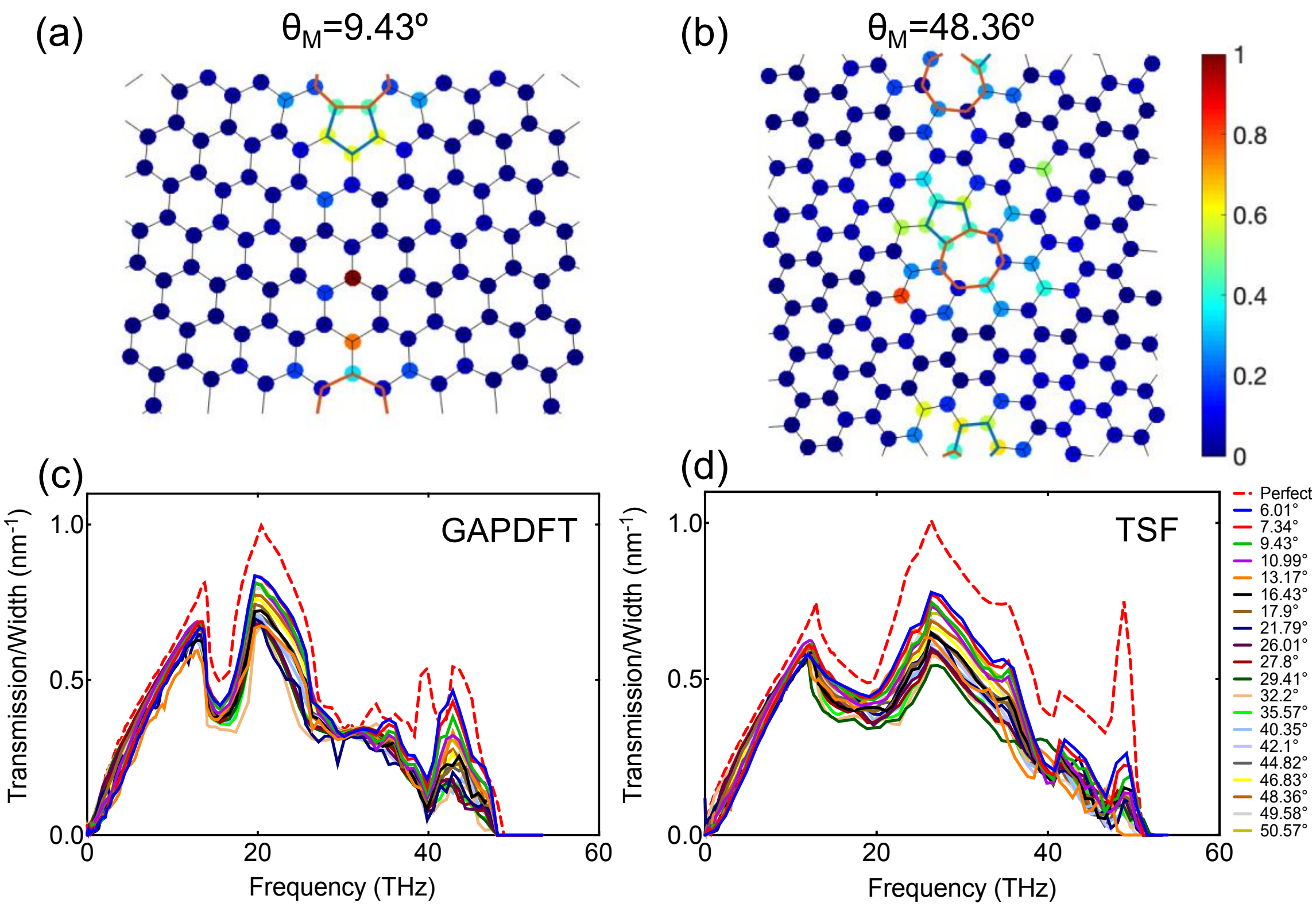

**Figure 7.** Comparison of TSF and GAPDFT showing inaccuracy of TSF for predicting force constants on dislocation cores and transmission function above mid phonon frequency. (a-b) normalized error of self-interatomic force constants, defined as $\left|\Delta\phi_{ii,\mathrm{TSF}} - \Delta\phi_{ii,\mathrm{GAPDFT}}\right| / \Delta\phi_{ii,\mathrm{GAPDFT}}$ where $\Delta\phi_{ii}$ is the difference of self-interaction force constants in GB and perfect graphene. (c-d) suppressed transmission function from perfect graphene for 20 GBs. The values in the legend are misorientation angles.

**Supplementary information**

# Ab initio phonon transport across grain boundaries in graphene using machine learning based on small dataset

Amirreza Hashemi[1], Ruiqiang Guo[2,#], Keivan Esfarjani[3,4,5], and Sangyeop Lee[2,6*]

[1] Computational Modeling and Simulation Program,
University of Pittsburgh, Pittsburgh PA 15261

[2] Department of Mechanical Engineering and Materials Science,
University of Pittsburgh, Pittsburgh PA 15261

[3] Department of Mechanical and Aerospace Engineering,
University of Virginia, Charlottesville VA 22904

[4] Department of Materials Science and Engineering,
University of Virginia, Charlottesville VA 22904

[5] Department of Physics,
University of Virginia, Charlottesville VA 22904

[6] Department of Physics and Astronomy,
University of Pittsburgh, Pittsburgh PA 15261

[#] current affiliation: Thermal Science Research Center, Shandong Institute of Advanced Technology, Jinan, Shandong Province, 250103, China.

* sylee@pitt.edu

### S1. Basic features of grain boundaries in graphene

Two dimensional materials have two degrees of freedom in the grain boundary (GB) configuration space shown in Figure S1. One is a misorientation angle ($\theta_M$) which is the smallest rotation angle making one grain aligned with the other grain, which is $\theta_1 + \theta_2$ in Figure S1. The $\theta_M$ ranges from 0° to 60°. The other is a line angle (LA), which measures the asymmetry of two grains about the GB that can be calculated as $\theta_1 - \theta_2$. In this work, we consider symmetric GBs only where the two grains are symmetric about the grain boundary and the line angle is zero. A previous study shows that characteristics of grain boundaries such as dislocation density, grain boundary formation energy, and the out-of-plane buckling do not depend on the line angle[1].

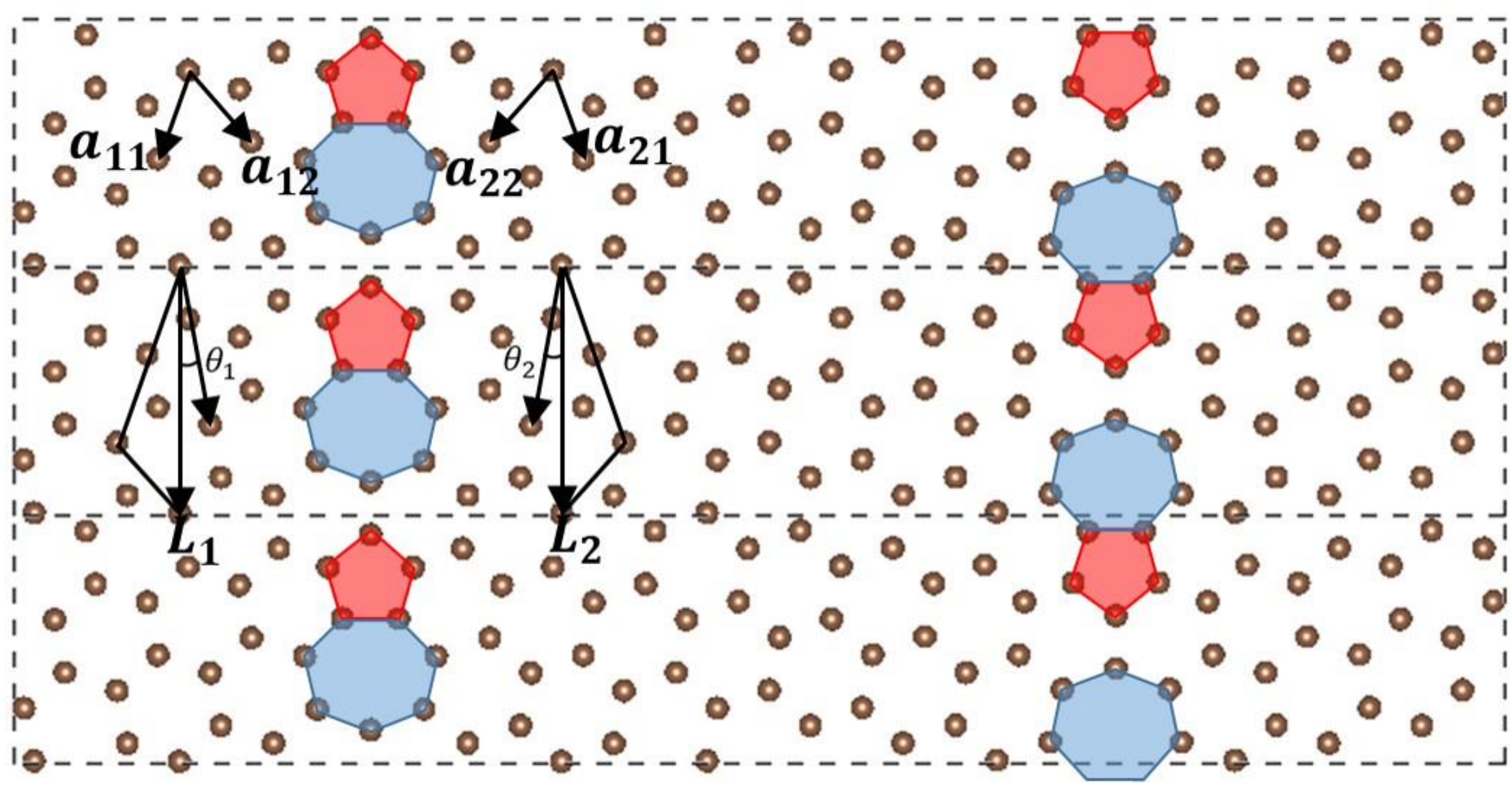

**Figure S1.** A typical structure containing two identical GBs with periodic boundary conditions along both the parallel and perpendicular directions to the GB. Each rectangle shown with dashed line represents one unit cell.

As it is shown in Figure S2, the graphene GBs show two combinations of primary disclinations (pentagon and heptagon rings), one is the GBs with the defects that are straight to the GB direction and has the Burger's vector of (1,0), and the second is the GBs with the defects that are not straight to the GB direction with the Burgers' vector of (1,0)-(0,1).

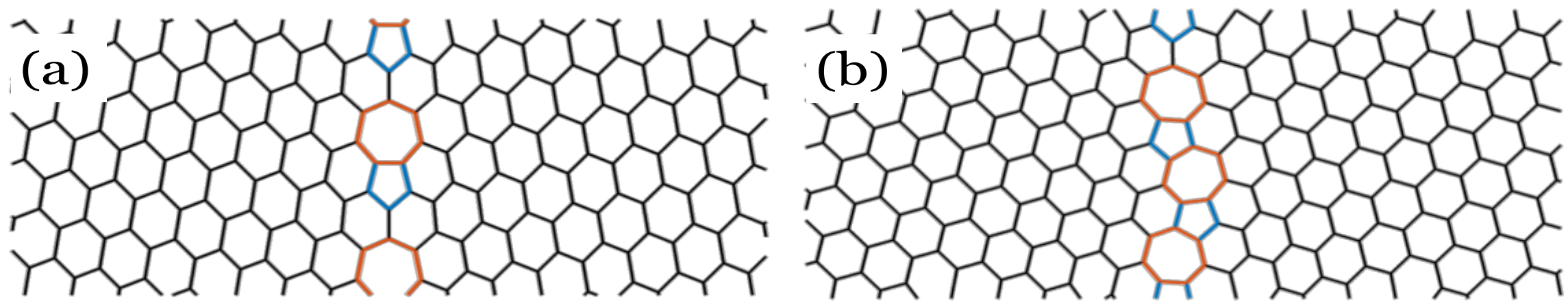

**Figure S2.** Two types of dislocation in graphene with the Burger's vector of (a) (1,0) and (b) (1,0)-(0,1)

It is useful to divide the entire range of $\theta_M$ into three groups to characterize the GBs: the low $\theta_M$ (<15°), mid $\theta_M$ (15°<MA<45°), and high $\theta_M$ (>45°). Low and high $\theta_M$ structures consist of (1,0) and (1,0)-(0,1) types, respectively, and have relatively low dislocation densities. However, GBs with mid $\theta_M$ have high dislocation density and the dominant type of defects changes from (1,0) to (1,0)-(0,1) as $\theta_M$ increases. Further detailed characteristics of GB structure with varying MAs was comprehensively discussed in previous studies[1].

In addition to disorder in the in-plane surface of graphene GBs, the out-of-plane buckling is common for most GB structures[1-3] which is a unique characteristic of 2D materials that 3D materials do not possess. The out-of-plane buckling reduces the local strain and stress and thus the GB structures with the lowest formation energy have the buckling.

## S2. Details of the chosen 20 GBs

In this study, we chose 20 GBs that covers the entire range of $\theta_M$. The following Table S1 summarize the characteristics including $\theta_M$ and the coincidence site lattice number (CSL, ∑). Note that the structures relaxed by the Tersoff (TSF) potential and density functional theory (DFT) calculation are slightly different and thus so are their GB period and disclination density.

**Table S1.** List of selected GBs with their structural properties

| index | $\theta_M$ (deg.) | CSL ∑ | Structures relaxed by TSF | | Structures relaxed by DFT | |
|---|---|---|---|---|---|---|
| | | | GB period (Å) | Disclination density ($Å^{-1}$) | GB period (Å) | Disclination density ($Å^{-1}$) |
| 1 | 6.01 | 91 | 23.7608 | 0.0842 | 23.3084 | 0.0858 |
| 2 | 7.34 | 61 | 19.4537 | 0.1028 | 19.0835 | 0.1048 |
| 3 | 9.43 | 37 | 15.1509 | 0.132 | 14.8627 | 0.1346 |
| 4 | 10.99 | 109 | 26.0123 | 0.1537 | 25.5172 | 0.1568 |
| 5 | 13.17 | 19 | 10.8593 | 0.1842 | 10.6527 | 0.1877 |
| 6 | 16.43 | 49 | 17.4449 | 0.2293 | 17.1116 | 0.2337 |
| 7 | 17.9 | 93 | 24.0326 | 0.2497 | 23.5727 | 0.2545 |
| 8 | 21.79 | 7 | 6.6012 | 0.3029 | 6.4725 | 0.309 |
| 9 | 26.01 | 79 | 22.1546 | 0.3611 | 21.7306 | 0.3681 |
| 10 | 27.8 | 39 | 15.5647 | 0.3855 | 15.2662 | 0.393 |
| 11 | 29.41 | 97 | 24.5397 | 0.4075 | 24.0737 | 0.4154 |
| 12 | 32.2 | 13 | 8.9919 | 0.4448 | 8.8178 | 0.4536 |
| 13 | 35.57 | 67 | 20.3977 | 0.3922 | 20.0072 | 0.3998 |
| 14 | 40.35 | 103 | 25.2933 | 0.3163 | 24.8106 | 0.3224 |
| 15 | 42.1 | 31 | 13.8792 | 0.2882 | 13.6149 | 0.2938 |
| 16 | 44.82 | 43 | 16.3393 | 0.2448 | 16.0284 | 0.2496 |
| 17 | 46.83 | 57 | 18.8093 | 0.2127 | 18.4503 | 0.2168 |
| 18 | 48.36 | 73 | 21.2859 | 0.1879 | 20.8804 | 0.1916 |
| 19 | 49.58 | 91 | 23.7631 | 0.1683 | 23.3122 | 0.1716 |
| 20 | 50.57 | 111 | 26.2369 | 0.1524 | 25.7374 | 0.1554 |

Figure S3 shows the density of disclination with respect to the $\theta_M$. The disclination density has a linear relation with $\theta_M$, having the maximum at the $\theta_M$ of around 35°.

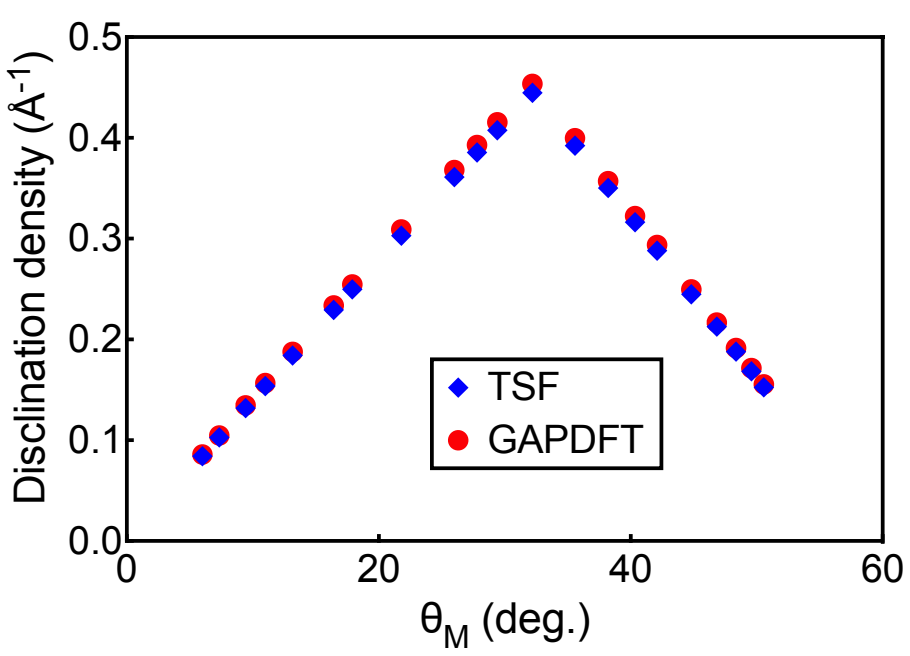

**Figure S3.** Density of disclination for selected 20 GBs relaxed with TSF and GAPDFT.

**S3. Generation of the atomic structures for training Gaussian approximation potential (GAP) and the atomistic Green's function (AGF) calculation.**

We generated supercells, each of which contain one GB from the 20 GB we chose, for GAP training and AGF calculation. The supercell for the AGF calculation is much larger than that for the GAP training. For the AGF calculation, the lead or contact serving as a heat reservoir needs to be crystalline graphene without strain induced by a GB. Thus, the supercells should be large enough so that a GB is placed far from the leads eliminating the strain in the lead. In our case, each supercell for the GAP training contain less than 500 atoms whereas the supercell for the AGF calculation is roughly 10 times larger than the supercell for the GAP training. The following Table S2 summarizes the size of supercells for the GAP training and AGF calculation.

For each supercell, we generated an initial atomistic structure using the code from previous work[1] based on the selected $\theta_M$ and the length of structure for one GB and then we inverted and merged the second inversely located GB to obtain the structure with two GBs and periodic boundary condition along the GB direction. Then, we relaxed the supercell using the TSF potential and DFT simulation with energy and force tolerances of 1e-6 eV and 1e-7 eV/Å respectively. The atomistic structures relaxed by the TSF and DFT are similar to each other, but slightly different as seen in Table S1. For the supercells that were used to generate training dataset of GAP, we amended the same supercell that is rotated by 180° to the original supercell so that it satisfies the translational invariance along the perpendicular direction to the GB. An example is shown in Figure S1.

**Table S2.** List of supercells for GAP training and AGF calculation. The length of supercell is the dimension along the heat flow direction, i.e., perpendicular to a GB. The length is from the structures relaxed by DFT

| index | $\theta_M$ (deg.) | Supercells for GAP | | Supercells for AGF | |
|---|---|---|---|---|---|
| | | # of atoms | Length (Å) | # of atoms | Length (Å) |
| 1 | 6.01 | 356 | 38.109 | 3676 | 407.494 |
| 2 | 7.34 | 276 | 35.996 | 2608 | 353.45 |
| 3 | 9.43 | 224 | 36.807 | 1500 | 261.129 |

| 4 | 10.99 | 368 | 35.342 | 4648 | 471.233 |
|---|---|---|---|---|---|
| 5 | 13.17 | 160 | 36.547 | 776 | 188.719 |
| 6 | 16.43 | 260 | 39.675 | 1996 | 301.466 |
| 7 | 17.9 | 336 | 35.584 | 1316 | 144.858 |
| 8 | 21.79 | 96 | 38.463 | 680 | 136.343 |
| 9 | 26.01 | 328 | 38.142 | 3232 | 385.202 |
| 10 | 27.8 | 216 | 35.997 | 554 | 94.007 |
| 11 | 29.41 | 344 | 38.23 | 4200 | 451.203 |
| 12 | 32.2 | 128 | 37.657 | 1320 | 388.38 |
| 13 | 35.57 | 300 | 37.851 | 2764 | 357.0618 |
| 14 | 40.35 | 372 | 36.804 | 4232 | 441.137 |
| 15 | 42.1 | 196 | 36.158 | 2616 | 497.056 |
| 16 | 44.82 | 240 | 37.323 | 3500 | 564.489 |
| 17 | 46.83 | 276 | 37.337 | 2684 | 376.326 |
| 18 | 48.36 | 324 | 37.8 | 3400 | 421.036 |
| 19 | 49.58 | 336 | 35.589 | 3884 | 430.688 |
| 20 | 50.57 | 408 | 39.606 | 1440 | 144.769 |

## S4. Details of training GAP

In the following table, we present the hyperparameters used for training GAPTSF and GAPDFT.

**Table S3.** List of hyperparameters for GAPTSF and GAPDFT

| Hyperparameter | Note | 2-body | 3-body | SOAP |
|---|---|---|---|---|
| $r_{cut}$ (Å) | Cutoff radius of the descriptor | 4.0 | 4.0 | 4.0 |
| $d$ (Å) | Transition width over which the magnitude of SOAP descriptor monotonically decrease to 0 | - | - | 1.0 |
| $\delta$ (eV) | Weight of different descriptors | 10.0 | 3.7 | 0.07 |
| $N_t$ | Number of representative atomic environments selected using the corresponding sparse method | 50 | 200 | 650 |
| Sparse method | | Uniform | Uniform | CUR |
| $n_{max}$ | Radial basis cutoff | - | - | 12 |
| $l_{max}$ | Angular basis cutoff | - | - | 12 |
| $\sigma_{energy}$ (eV/atom) | Expected error for atomic energy | 0.001 | | |
| $\sigma_{force}$ (eV/Å) | Expected error for force | 0.0005 | | |

## S5. Self-interaction force constants from TSF and GAPDFT

Here we present the magnitude of the self-interaction force constants for the comparison of TSF and GAPDFT potentials. The results in Figure S4 show that the difference between TSF and GAPDFT as large as 35 %. We observed that the differences between the interacting force constants are consistent between GAPDFT and TSF for all the GBs including training and testing GB structures.

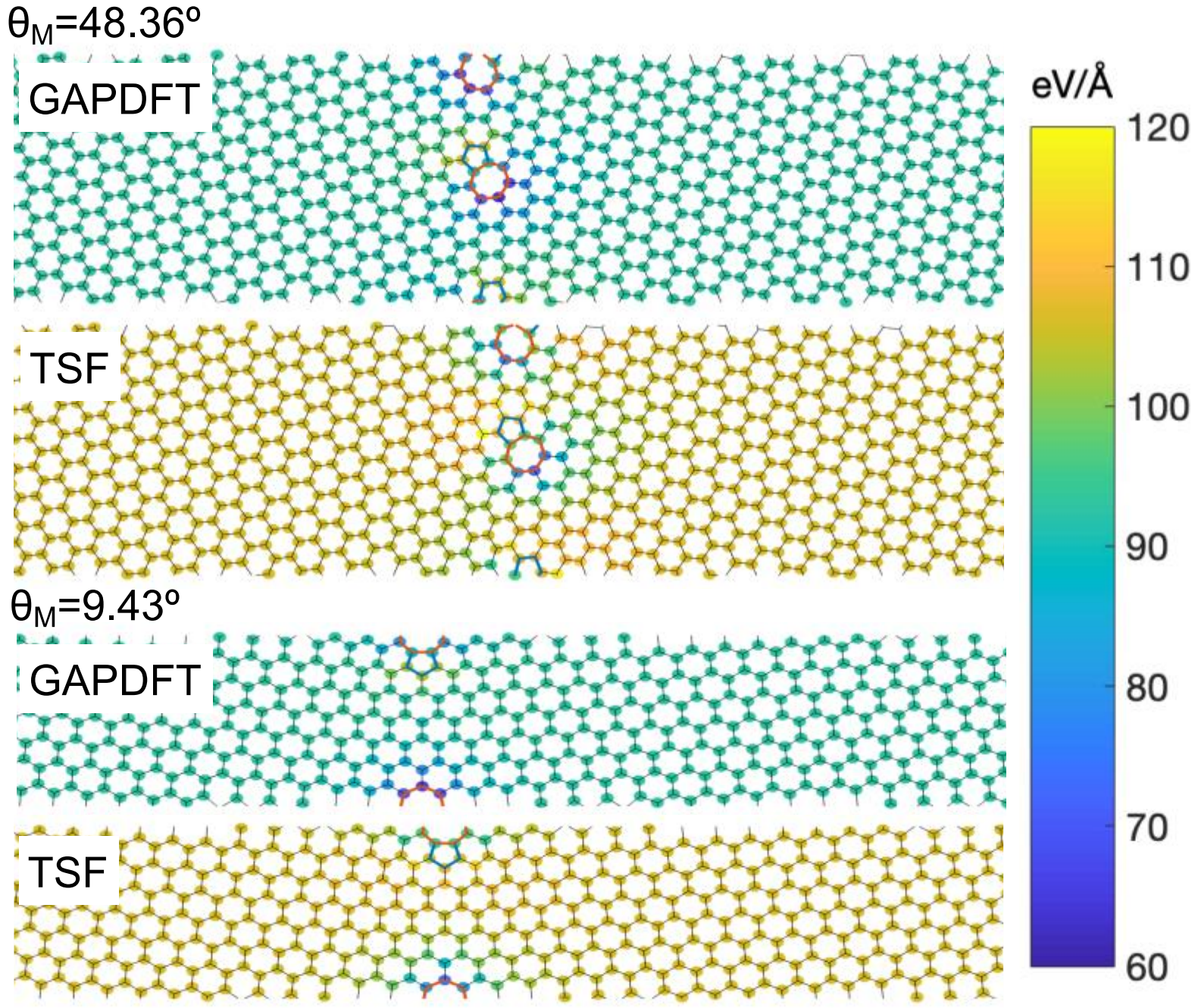

**Figure S4.** Comparison of magnitudes of self-interaction force constants from GAPDFT and TSF for a training structure with $\theta_M$=48.36° and a test structure with $\theta_M$=9.43°